\documentclass[aps,prd,amsmath
,eqsecnum            
,preprintnumbers
,nofootinbib         
 ,showpacs          
 ,showkeys          
,twocolumn         
]{revtex4}
\usepackage[dvips]{graphicx}
\usepackage{amsmath}
\usepackage{amsfonts}
\usepackage{amssymb}
\usepackage{longtable}   
\usepackage{color}

\allowdisplaybreaks[4]     

\newcommand{\df}{\ {\overset {\rm def} =}\ }
\newcommand{\dr}[2]{\frac {{\rm d} {#1}} {{\rm d} {#2}}}

\newcommand{\dril}[2]{{{\rm d} {#1}} / {{\rm d} {#2}}}

\begin{document}

\title{Gamma ray bursts may be blueshifted bundles of the relic radiation}

\author{Andrzej Krasi\'nski}
\affiliation{N. Copernicus Astronomical Centre, Polish Academy of Sciences, \\
Bartycka 18, 00 716 Warszawa, Poland} \email{akr@camk.edu.pl}

\date {}

\begin{abstract}
A hypothesis is proposed that the gamma-ray bursts (GRBs) may arise by
blueshifting the emission radiation of hydrogen and helium generated during the
last scattering epoch. The blueshift mechanism is provided by such a
Lema\^{\i}tre -- Tolman (L--T) model, in which the bang-time function $t_B(r)$
is not everywhere constant. Blueshift arises on \textit{radial} rays that are
emitted over regions where $\dril{t_B} r \neq 0$. The paper presents an L--T
model adapted for this purpose and shows how it accounts for the observed
properties of the GRBs; some properties are accounted for only qualitatively.
\end{abstract}

\maketitle

\section{Motivation and background}\label{intro}

\setcounter{equation}{0}

In the Lema\^{\i}tre \cite{Lema1933} -- Tolman \cite{Tolm1934} (L--T) models, in
which the bang-time function $t_B(r)$ is not everywhere constant,
\textit{radial} light rays emitted close to those points of the Big Bang (BB) at
which $\dril {t_B} r \neq 0$ display blueshifts to later observers (the
blueshift is infinite, $z = -1$, on rays emitted exactly at the BB
\cite{Szek1980,HeLa1984,Kras2014d}).

On the other hand, gamma-ray bursts (GRBs) are observed and are believed to
originate at large distances from our Galaxy, up to several billion light years
\cite{gammainfo}. The question thus arises: could GRBs have been emitted in the
last scattering epoch, together with the relic radiation now observed as the
cosmic microwave background (CMB), and then blueshifted to gamma-ray frequencies
by the mechanism mentioned above?

For the blueshift mechanism to work, the GRBs would have to originate in regions
that emerged from a locally delayed BB.\footnote{Regions where the BB occurred
\textit{earlier} than in the background generate shell crossing singularities
\cite{Szek1980,PlKr2006}.} The relic radiation is emitted a finite time after
the BB, so its blueshift must be bounded from below ($z \geq z_{\rm min} > -1$).
The technical problem to solve is this: Is $z_{\rm min}$ sufficiently small
that, with the free functions of the L--T model suitably chosen, the frequencies
are blueshifted from the range of the emission spectra of hydrogen and helium
(the only elements present in large amounts during last scattering) to the
gamma-ray range observed today? The present paper attempts to answer this
question in the positive -- see Sec. \ref{modelfit}.

Section \ref{GRBdata} provides the most basic information on the GRBs. Section
\ref{LTintro} is an introduction to the L--T models, and Sec. \ref{LTnullgeo}
provides information on light propagation in these models. Section
\ref{modelfit} presents the current best-fit L--T model that reflects the
properties of the GRBs, and discusses improvements in the model needed to
achieve a full quantitative fit. Conclusions are summarized in Sec.
\ref{conclu}. The Appendix presents the data needed to replicate the numerical
calculations.

\section{Basic facts about the GRBs}\label{GRBdata}

\setcounter{equation}{0}

The following properties of the GRBs need to be accounted for \cite{gammainfo}:

(1) Their frequencies extend from $\nu_{\gamma {\rm min}} \approx 0.24 \times
10^{19}$ Hz to $\nu_{\gamma {\rm max}} \approx 1.25 \times 10^{23}$ Hz
\cite{Gold2012}\footnote{Converted from keV to Hz by: $1{\rm eV} = 1.6 \times
10^{-19} {\rm J} = {\rm h} \times 0.24 \times 10^{15}$ Hz, where h = $6.626
\times 10^{-34}$ J s \cite{unitconver,constants}.} (see also Ref.
\cite{Gold2014}).

(2) They last from less than a second to a few minutes.

(3) They are probably focussed into narrow jets.

(4) A GRB is sometimes followed by a longer-lived and fainter ``afterglow'' at
larger wavelengths.

(5) Nearly all GRBs come from very large distances, from over $10^8$ to several
billion light years.

Currently, there is no generally accepted explanation of origins of the GRBs.
There exist only attempts at explanation by known astrophysical phenomena such
as gravitational collapse to a black hole, a supernova explosion or a collision
of ultra-dense neutron stars \cite{gammainfo}.

The model presented in Sec. \ref{modelfit} accounts quantitatively for the lower
limit in property (1) and for (5), qualitatively for (3-4), and is not in
contradiction with (2). References to these properties will be marked there by
bullets \textcolor[rgb]{1.00,0.00,0.50}{{\Huge {$\bullet$}}}. To achieve a
quantitative agreement, more elaborate fitting will be needed.

\section{The L--T models}\label{LTintro}

\setcounter{equation}{0}

The metric of the L--T models is:
\begin{equation}\label{3.1}
{\rm d} s^2 = {\rm d} t^2 - \frac {{R_{,r}}^2}{1 + 2E(r)}{\rm d} r^2 -
R^2(t,r)({\rm d}\vartheta^2 + \sin^2\vartheta \, {\rm d}\varphi^2),
\end{equation}
where $E(r)$ is an arbitrary function. The source in the Einstein equations is
dust; its (geodesic) velocity field is
\begin{equation}\label{3.2}
u^{\alpha} = {\delta^{\alpha}}_0.
\end{equation}
Because of the assumption $p = 0$ built into this model, it is inadequate before
the last-scattering epoch.

The function $R(t, r)$ is determined by
\begin{equation}\label{3.3}
{R_{,t}}^2 = 2E(r) + 2M(r) / R,
\end{equation}
$M(r)$ being another arbitrary function; we neglect the cosmological constant.
We will consider only the models with $R,_t > 0$ and $E > 0$. The solution of
(\ref{3.3}) is then:
\begin{eqnarray}\label{3.4}
R(t,r) &=& \frac M {2E} (\cosh \eta - 1), \nonumber \\
\sinh \eta - \eta &=& \frac {(2E)^{3/2}} M \left[t - t_B(r)\right],
\end{eqnarray}
where $t_B(r)$ is one more arbitrary function; the BB occurs at $t = t_B(r)$.
The mass density is
\begin{equation}  \label{3.5}
\kappa \rho = \frac {2{M_{,r}}}{R^2R_{,r}}, \qquad \kappa \df \frac {8\pi G}
{c^2}.
\end{equation}
The $r$-coordinate is chosen so that \cite{Kras2014d}
\begin{equation}\label{3.6}
M = M_0 r^3,
\end{equation}
and $M_0 = 1$ (kept in formulae for dimensional clarity).

The units used in numerical calculations were introduced and justified in Ref.
\cite{Kras2014}. Taking \cite{unitconver}
\begin{equation}\label{3.7}
1\ {\rm pc} = 3.086 \times 10^{13}\ {\rm km}, \quad 1\ {\rm y} = 3.156 \times
10^7\ {\rm s},
\end{equation}
the numerical length unit (NLU) and the numerical time unit (NTU) are defined as
follows:
\begin{equation}\label{3.8}
1\ {\rm NTU} = 1\ {\rm NLU} = 9.8 \times 10^{10}\ {\rm y}.
\end{equation}

\section{Light rays in an L--T model}\label{LTnullgeo}

\setcounter{equation}{0}

The geodesic null vector fields $k^{\alpha}$ in (\ref{3.1}) obey \cite{BKHC2010}
\begin{equation} \label{4.1}
\left(k^t\right)^2 - \frac {{R,_r}^2 \left(k^r\right)^2} {1 + 2E} - \frac {C^2}
{R^2} = 0,
\end{equation}
with $C$ being constant along a geodesic.

The general formula for redshift is \cite{Elli1971}
\begin{equation}\label{4.2}
1  + z = \frac {(u_{\mu} k^{\mu})_e} {(u_{\nu} k^{\nu})_o},
\end{equation}
where $k^{\mu}$ is an affinely parametrised vector field tangent to a light ray
connecting the source and the observer, both comoving with the cosmic medium
with the four-velocity $u^{\alpha}$. The subscript ``{\it e}'' means ``at the
emission event'', ``{\it o}'' means ``at the observation event''. We will
consider\ \ past-directed\ \ rays,\ \ on which\ \ $k^t < 0$. We can

\noindent rescale the affine parameter $\lambda$ so that
\begin{equation}
k^t(t_o) = -1. \label{4.3}
\end{equation}
Then, with $u^{\alpha}$ being given by (\ref{3.2}), we have
\begin{equation}\label{4.4}
1 + z = - k^t.
\end{equation}
For nonradial rays, on which $C \neq 0$, the last term in (\ref{4.1}) will go to
infinity when $R \to 0$. Thus, at the BB
\begin{equation}\label{4.5}
\lim_{R \to 0} \left|k^t\right| = \infty.
\end{equation}
Equations (\ref{4.5}) and (\ref{4.4}) imply that $z \to \infty$ at the BB on all
nonradial rays, in agreement with Ref. \cite{HeLa1984}.

For radial rays, $z \to -1$ as $R \to 0$ when $\dril {t_B} r \neq 0$ at the
intersection of the ray with the BB, and $z \to \infty$ when $\dril {t_B} r = 0$
\cite{HeLa1984,Kras2014d} . The result $z = -1$ implies infinite blueshift for
all observers. Rays emitted close to, but not right at the BB will acquire a
finite blueshift that can be overcompensated by later-acquired redshifts if the
observation is carried out sufficiently far to the future from the emission
point.

In consequence of (\ref{4.5}), blueshifts may arise \textit{only on radial
rays}. Thus, for an observer re-directing her telescope away from the direction
to the center of the radiation source, the transition from blueshift to redshift
would occur abruptly -- if the real Universe were exactly modelled by the L--T
geometry. In reality, the changeover from blueshift to redshift can be expected
to occur in a finite, but short time. This would account for the short-livedness
of the GRBs (property (2) in Sec. \ref{GRBdata}), their narrow-jet appearance
(property (3)) and for their ``afterglows'' (property (4)) -- see Sec.
\ref{modelfit} for details.

In the following, past-directed radial rays are dealt with, on which $C = 0$.
Using (\ref{4.1}), the equations to be integrated numerically are:
\begin{eqnarray}
\dr t {\lambda} &=& k^t, \qquad \dr {k^t} {\lambda} = - \left(k^t\right)^2 \frac
{R,_{t r}} {R,_r}, \label{4.6} \\
k^r &=& \pm \frac {\sqrt{1 + 2E}} {R,_r}\ k^t, \qquad \dr r {\lambda} = k^r,
\label{4.7}
\end{eqnarray}
with the initial condition (\ref{4.3}). The sign in (\ref{4.7}) is $+$ on
past-inward rays and $-$ on past-outward rays.

In a general L--T model with $E \neq 0$ we have \cite{PlKr2006}
\begin{eqnarray}
R,_r&=& \left(\frac {M,_r} M - \frac {E,_r} E\right)R + \Psi(t, r) R,_t,
\label{4.8} \\
R,_{tr} &=& \frac {E,_r} {2E}\ R,_t - \frac M {R^2}\ \Psi(t, r),
\label{4.9} \\
\Psi(t, r) &\df& \left(\frac 3 2 \frac {E,_r} E - \frac {M,_r} M\right)
\left(t - t_B\right) - t_{B,r}. \ \ \ \ \ \label{4.10}
\end{eqnarray}
As can be seen from (\ref{4.6}) and (\ref{4.4}), $R,_{tr} = 0$ is the locus of
extrema of redshift along radial rays; we call it the maximum-redshift
hypersurface (MRH).

Consider a radial ray proceeding to the past from an initial point that lies
later than the MRH. The redshift $z$ on it increases from 0 to a maximum,
achieved at the MRH. Further down the ray, $z$ decreases. If the ray could
continue to the BB, $z$ would decrease to $-1$. However, the L--T model ceases
to apply at the last-scattering hypersurface (LSH). Can $z$ become sufficiently
negative before the ray crosses the LSH for shifting the optical frequencies to
the gamma-ray range? It is shown in Sec. \ref{modelfit} that this is indeed
possible when the functions $E(r)$ and $t_B(r)$ are suitably chosen, and the
observer is put in the right place at the right time.

\section{Fitting the L--T model to observations and
measurements}\label{modelfit}

\setcounter{equation}{0}

The mass density at the instant of last scattering in the standard $\Lambda$CDM
model is \cite{Kras2014c}
\begin{equation}\label{5.1}
\kappa \rho_{\rm ls} \approx 88 \times 10^9\ ({\rm NLU})^{-2}.
\end{equation}
We assume that the recombination occurs at the same density also in an
inhomogeneous Universe. The density along a ray is calculated using (\ref{3.5}),
and the value of $z$ at the moment when $\rho = \rho_{\rm ls}$ emerges from
(\ref{4.4}) and (\ref{4.6}).

The GRBs cannot arise by blueshifting the whole black-body spectrum of the relic
radiation to the gamma-ray range. First, the spectra of the GRBs do not have the
black-body forms (example: Ref. \cite{GRBrealspectra}). Second, the intensity of
a GRB created in this way would exceed the observed ones by tens of orders of
magnitude.\footnote{The Planck formula and the current CMB spectrum
\cite{CMBspectrum} imply $1.295 \times 10^9$ W/(cm$^2 \times$ sr $\times$ Hz)
for the maximum intensity of the black-body spectrum blueshifted from LSH by $1
+ z = 10^{-5}$. Observed GRBs have intensities below $10^{-24}$ W/(cm$^2$ Hz)
\cite{McBr2006}.} So, if the GRBs arise by blueshifting the relic radiation,
then the different frequencies have to be blueshifted individually.

To shift the lowest frequency of the hydrogen emission spectrum
\cite{Hydrospec}, $\nu_{\rm Hmin} = 4.054 \times 10^{13}$ Hz (corresponding to
the wavelength of 7400 nm) to the lowest observed frequency of the gamma-ray
bursts, $\nu_{\gamma {\rm min}} \approx 0.24 \times 10^{19}$ Hz \cite{Gold2012},
the blueshift $1 + z \approx 1.667 \times 10^{-5}$ is needed. To shift the
frequency of the most intense line in the hydrogen spectrum, $2.1876 \times
10^{15}$ Hz (corresponding to $656.2852$ nm) to the same minimum gamma-ray
frequency, $1 + z \approx 1.9 \times 10^{-4}$ is needed. To shift the maximum
measured hydrogen emission frequency, $\nu_{\rm Hmax} = 3.2 \times 10^{15}$ Hz
(corresponding to 93.782 nm) to the maximum recorded cosmic gamma-ray frequency,
$\nu_{\gamma {\rm max}} \approx 1.25 \times 10^{23}$ Hz \cite{Gold2012}, $1 + z
\approx 2.56 \times 10^{-8}$ is needed.\footnote{The most intense helium
emission lines have wavelengths between 388 nm and 846 nm, i.e. within the range
of the hydrogen spectrum \cite{helspectr}.}

\begin{figure}[h]
\hspace{-4mm}
\includegraphics[scale=0.5]{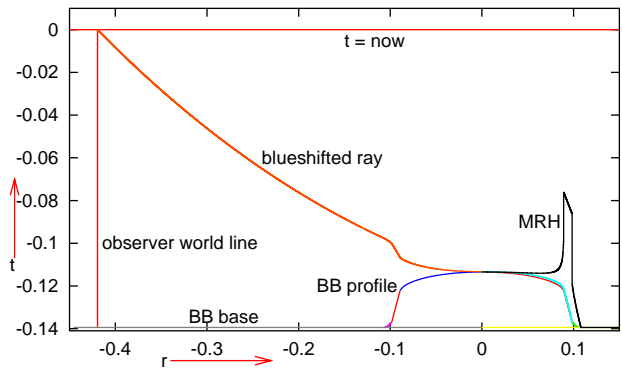}
\caption{The blueshifted ray, the bang-time profile and the profile of the
maximum-redshift hypersurface in the L--T model described in the text. See the
Appendix for the parameters of $t_B(r)$. The MRH is symmetric around the $r = 0$
line, but its left part is suppressed in this graph.} \label{drawtrueray}
\end{figure}

The configuration shown in Fig. \ref{drawtrueray} was obtained by fitting the
functions $E(r)$ and $t_B(r)$ and the position of the observer by trial and
error so as to make $1 + z$ as close to zero as possible. The current best
result is
\begin{equation}\label{5.2}
1 + z_{\rm mb} = 1.23007568 \times 10^{-5}
\end{equation}
between the LSH and now. This accounts for the lower end of frequencies in
\textcolor[rgb]{1.00,0.00,0.50}{{\Huge {$\bullet$}} Property (1)}. This is the
minimum $z$ within the chosen class of BB profiles; other profile classes may
possibly lead to smaller $z$.

For simplicity, $E$ was chosen the same as in a Friedmann model, with
\begin{eqnarray}\label{5.3}
2E/r^2 \df -k = - 0.4.
\end{eqnarray}
A general $E$ would have $-k$ replaced by $(-k + {\cal F}(r))$, where ${\cal
F}(0) = 0$, but otherwise is arbitrary \cite{PlKr2006}. This would provide
further parameters for fine-tuning.

The BB profile consists of a spherically symmetric hump surrounding the center
of symmetry out to a finite distance; further away from the center $t_B$ is
constant, and so the geometry is Friedmannian. The present time is $t = 0$, and
the flat part of $t_B$ was chosen at
\begin{eqnarray}\label{5.4}
t \df t_{\rm Bf} &=& -0.13945554689046649\ {\rm NTU} \nonumber \\
&\approx& -13.67 \times 10^9\ {\rm years};
\end{eqnarray}
this is the asymptotic value of $t_B$ in the L--T model that imitates
accelerated expansion using nonconstant $t_B$ \cite{Kras2014d}.

A radial cross-section through the hump in the BB is shown in Fig.
\ref{drawfarkantray}. It consists of two ellipse arcs (see Fig.
\ref{drawpicture} in the Appendix) that are connected by a straight line segment
in the neigbourhood of the point, where the full ellipses would be tangent to
each other. The lower ellipse arc is tangent to the flat part of the $t_B$
profile. This shape is determined by five parameters: four semi-axes of the two
ellipses, and the tilt of the straight segment (see the Appendix for their
values).

\begin{figure}[h]
\hspace{-6mm}
\includegraphics[scale=0.5]{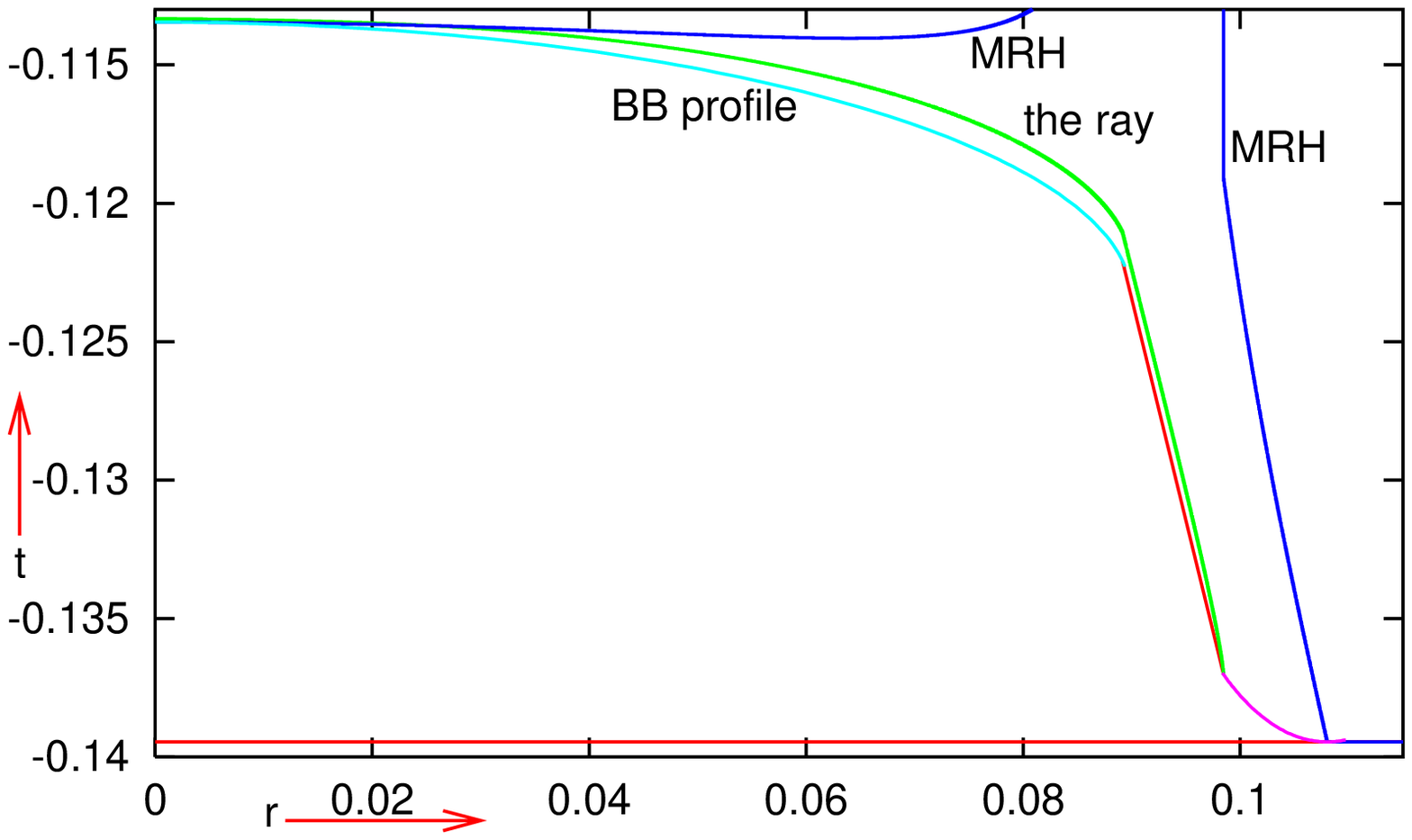}
${ }$ \\[-4cm]
\hspace{-1.8cm}
\includegraphics[scale = 0.4]{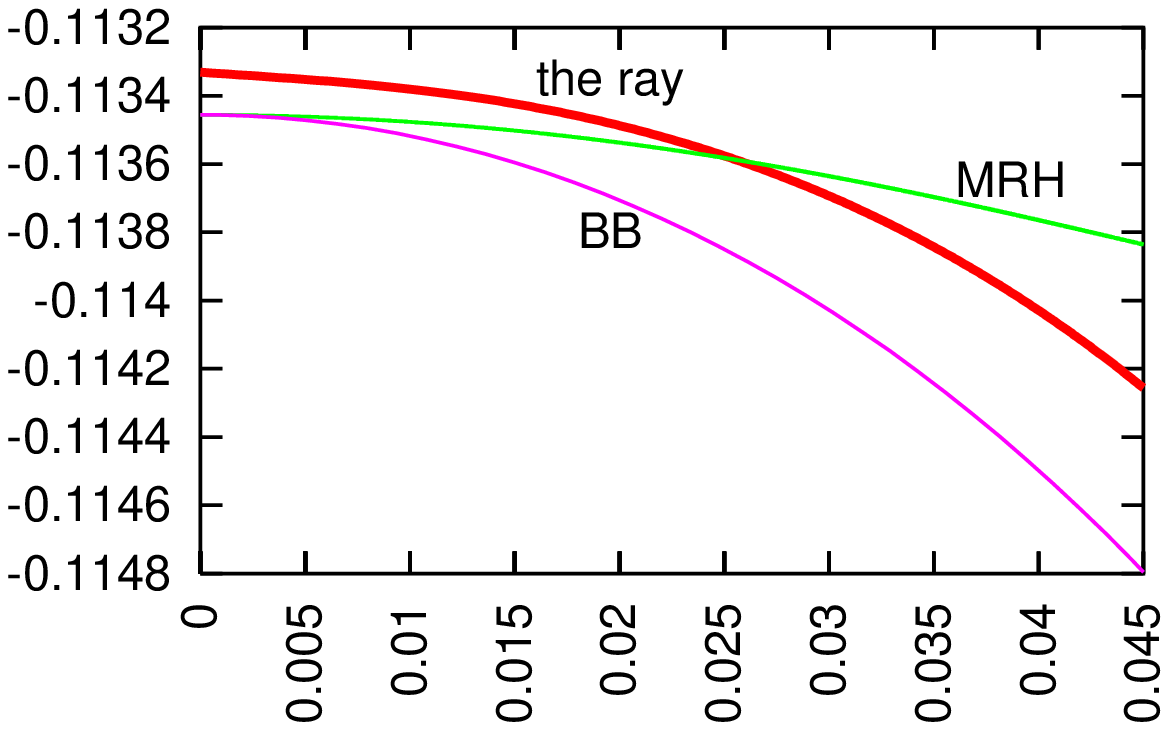}
\vspace{1cm} \caption{{\bf Main panel:} Closeup view of the neighbourhood of the
BB in Fig. \ref{drawtrueray}. {\bf Inset:} Closeup view of the neighbourhood of
the BB maximum -- see text for a comment.} \label{drawfarkantray}
\end{figure}

The inset in Fig. \ref{drawfarkantray} shows the neighbourhood of the maximum of
$t_B(r)$. The ray passes $\Delta t_c = 0.125 \times 10^{-3}$ NTU $= 1.225 \times
10^7$ years over this maximum. The $\Delta t_c$ is also an adjustable parameter.
The blueshift at the observer position is sensitive to the value of $\Delta t_c$
in certain ranges. For example, $\Delta t_c = 0.126395 \times 10^{-3}$ NTU
results in $1 + z \approx 3.9 \times 10^{-5}$, while $\Delta t_c = 0.1264 \times
10^{-3}$ NTU results in $1 + z \approx 4.5 \times 10^{-2}$. Thus, a change in
$\Delta t_c$ by $5 \times 10^{-9}$ NTU $\approx 49$ years causes that $1 + z$
goes up by the factor $0.87 \times 10^3$ -- enough for the observed radiation to
drift from the gamma-range into the ultraviolet \cite{elecspectr}. The 49 years
is not a sufficiently short time to account for the abrupt end of a GRB
quantitatively,\footnote{Moreover, the accuracy of numerical calculations for
this paper was insufficient to capture time intervals of the order of hours.}
but the qualitative effect of afterglow is here, and thereby
\textcolor[rgb]{1.00,0.00,0.50}{{\Huge {$\bullet$}} Property (4)} is
qualitatively accounted for.

But the abrupt beginning and end of a GRB and the afterglow should rather be
associated with the center of the radiation source going into and off the line
of sight than with the passage of time at the observer (see Sec.
\ref{LTnullgeo}). The model predicts a discontinuous jump from redshift to
blueshift and back to redshift for an arbitrarily small change in the direction
of observation (which qualitatively accounts for
\textcolor[rgb]{1.00,0.00,0.50}{{\Huge {$\bullet$}} Property (3)}). In reality
such a change might be provided, for example, by the orbital motion of the
Earth. Then, the blueshifted ray would stay within the observer's field of view
only briefly, and this would implicitly account for
\textcolor[rgb]{1.00,0.00,0.50}{{\Huge {$\bullet$}} Property (2)}.

If the center of the BB hump would stay in the observer's line of sight all the
time, then she would see the gamma radiation persisting for nearly the whole
$\Delta t_c = 1.225 \times 10^7$ years. (But this is a property only of the
concrete BB profile of Fig. \ref{drawfarkantray}. This does not exclude the
existence of profiles providing shorter viewing times.) It would show up
abruptly, because the ray emitted near the top of the hump would have large {\it
red}shift.

The model of Fig. \ref{drawtrueray} cannot explain the highest-frequency end of
the GRB spectrum \cite{Gold2012} by means of blueshifting from the hydrogen
emission range. This does not yet imply that the high-frequency GRBs must arise
by a different mechanism: it is still an open question whether a finer-tuned
L--T model could do the job.

A cosmological model that would account for the multitude of observed GRBs
should be imagined as a Friedmann background containing many humps like the one
in Fig. \ref{drawfarkantray}, of different shapes, different spatial extents and
different heights above the flat part of $t_B(r)$, placed at different comoving
positions.

Redshift is used in astronomy as the measure of distance. But blueshifting
renders redshift -- distance relations multivalued \cite{Kras2014c}. No
operational method of determining the distance is known when blueshifts are
present. The distance from the observer to the source of the ray in Fig.
\ref{drawtrueray} can be estimated in two ways:

1. The intersection of the ray with the LSH occurs $\approx 13.3764 \times 10^9$
years ago, so by conventional accounting the source would be $13.3764 \times
10^9$ light years from the observer. This is $6.83 \times 10^8$ years later than
the BB of the $\Lambda$CDM model, whose present age is \cite{Plan2013}
\begin{equation}\label{5.5}
T = 13.819 \times 10^9\ {\rm y} = 0.141\ {\rm NTU}.
\end{equation}

2. The sum of the values of the $r$-coordinate of the observer ($r_{\rm O} =
0.41946$) and of the intersection of the ray with the LSH ($r_{\rm LSHi} =
0.09841$) is $r_e = 0.51787$. The $r_e$ determines the active gravitational mass
contained within the $r = r_e$ sphere centered at the observer by $M = M_0
{r_e}^3$. In the $\Lambda$CDM model, the same $r_e$ (corresponding to the same
mass) is reached by the present observer's past light cone at $t = t_e =
-0.1279$ NTU, i.e. $\approx 1.25 \times 10^9$ years ago,\footnote{This
calculation was based on the numerical results of Ref. \cite{Kras2014b}.} which
implies the distance $\approx 1.25 \times 10^9$ light years. This number is
within the range of distances inferred from observations of the GRBs
\cite{gammainfo}, and thereby \textcolor[rgb]{1.00,0.00,0.50}{{\Huge
{$\bullet$}} Property (5)} is accounted for.

The most obvious possibility to improve the model is to manipulate the shape and
size of the BB hump seen in Fig. \ref{drawtrueray}, for example, by putting more
parameters into it. This may result in the rays staying before the MRH for a
longer segment of the affine parameter, and thus in smaller observed values of
$1 + z$.

The second possibility is to model the BB hump by the Szekeres metric
\cite{Szek1975,Szek1975b}, \cite{PlKr2006}. It contains the L--T model as a
subcase, but in general has no symmetry. Therefore, it must be verified what
directions in the Szekeres class of metrics go over into the L--T radial
directions, and whether blueshifts appear on them. A Szekeres model might
produce a different time-profile of the observed frequency, and a better
quantitative agreement with the observed properties of the GRBs.

\section{Conclusions}\label{conclu}

\setcounter{equation}{0}

The model proposed here deals satisfactorily with the low end of the GRB
frequency range (property (1) in Sec. \ref{GRBdata}) and with the low end of the
distance range (property (5)). It also deals qualitatively with poperties (3)
and (4). Re (3): an arbitrarily small misalignment between the line of sight and
the direction to the center of the radiation source causes the gamma-ray impulse
to disappear. Re (4): the afterglow is present in the model of Sec.
\ref{modelfit}, but its duration does not agree with observations. The only
property not explicitly accounted for is (2). In order to account for it, one
should calculate the change of observed frequency induced by a small change in
the direction of observation. This is possible to do in the present model, but
requires numerical calculations of a much higher precision.

Using a BB profile with more parameters, and modelling the BB hump using a
Szekeres rather than the L--T metric may lead to a full quantitative agreement
between the model and the observations.

\appendix

\section{The BB profile}\label{techdata}

\setcounter{equation}{0}

\begin{figure}[h]
\includegraphics[scale=0.5]{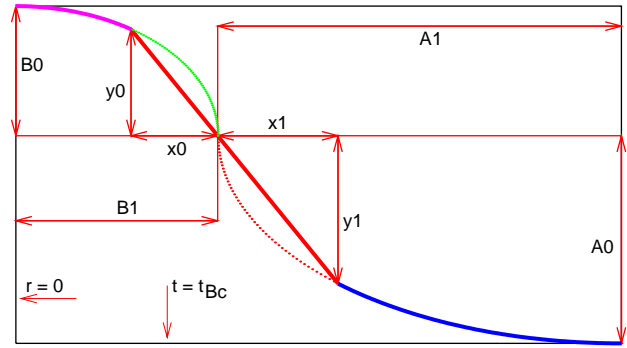}
\caption{Parameters of the bang-time profile (drawn not to scale for better
readability). See text for the actual values.} \label{drawpicture}
\end{figure}

The hump in the BB profile consists of two ellipse arcs and of a straight line
segment joining them that passes through the point where the full ellipses would
touch each other -- see Fig. \ref{drawpicture}. The values of the parameters in
Fig. \ref{drawfarkantray} are $B_0 = 0.01$, $B_1 = 0.09$, $A_0 = 0.016$, $A_1 =
0.018$, $x_0 = 0.000859$. The other ones are uniquely determined by these. The
units for $A_0$ and $B_0$ are NTU; the other parameters are dimensionless. The
values of $A_0$ and $B_0$ imply for the time difference between the maximum of
$t_B$ and its flat part 0.026 NTU = $2.548 \times 10^9$ years (0.18 of the age
of the Universe given by (\ref{5.5})).

\end{document}